\documentclass[conference]{IEEEtran}

\DeclareMathAlphabet{\mathpzc}{OT1}{pzc}{m}{it}

\newtheorem{lemma}{Lemma}

\newcommand{\qed}{\nobreak \ifvmode \relax \else
      \ifdim\lastskip<1.5em \hskip-\lastskip
      \hskip1.5em plus0em minus0.5em \fi \nobreak
      \vrule height0.75em width0.5em depth0.25em\fi}

\usepackage{psfrag, epsfig, amsfonts, algorithm, algorithmic}
\usepackage{cite, graphicx, psfrag, subfigure, url, stfloats, amsmath, amssymb}
\usepackage{graphicx}
\usepackage{epstopdf}
\begin{document}
%
\title{Wideband Spectrum Sensing in Cognitive Radio Networks} %

\author{\authorblockN{Zhi Quan$^\dag$, Shuguang Cui$^\ddag$, Ali H. Sayed$^\dag$, and H. Vincent
Poor$^\S$}
\authorblockA{$^\dag$ Department of Electrical Engineering, University of
California, Los Angeles, CA 90095  \\ $^\ddag$Department of
Electrical and Computer Engineering,
Texas A\&M University, College Station, TX 77843\\
$^\S$Department of Electrical Engineering, Princeton University, Princeton, NJ 08544\\
Email: \{quan, sayed\}@ee.ucla.edu; cui@ece.tamu.edu;
poor@princeton.edu }}
\markboth{}{}

\maketitle

\begin{abstract}

Spectrum sensing is an essential enabling functionality for
cognitive radio networks to detect spectrum holes and
opportunistically use the under-utilized frequency bands without
causing harmful interference to legacy networks. This paper
introduces a novel wideband spectrum sensing technique, called
\emph{multiband joint detection}, which jointly detects the signal
energy levels over multiple frequency bands rather than consider
one band at a time. The proposed strategy is efficient in
improving the dynamic spectrum utilization and reducing
interference to the primary users. The spectrum sensing problem is
formulated as a class of optimization problems in interference
limited cognitive radio networks. By exploiting the hidden
convexity in the seemingly non-convex problem formulations,
optimal solutions for multiband joint detection are obtained under
practical conditions. Simulation results show that the proposed
spectrum sensing schemes can considerably improve the system
performance. This paper establishes important principles for the
design of wideband spectrum sensing algorithms in cognitive radio
networks.

\end{abstract}


%
\IEEEpeerreviewmaketitle


\section{Introduction}

%

Spectrum sensing is an essential functionality of cognitive radios
since the devices need to reliably detect weak primary signals of
possibly-unknown types \cite{Cabric04}. In general, spectrum
sensing techniques can be classified into three categories: energy
detection \cite{Kay98II}, matched filter coherent detection
\cite{Poor1994}, and cyclostationary feature detection
\cite{Enserink1994}. Since non-coherent energy detection is simple
and is able to locate spectrum-occupancy information quickly, we
will adopt it as a building block for constructing the proposed
wideband spectrum sensing scheme.

There are previous studies on spectrum sensing in cognitive radio
networks with focus on cooperation among multiple cognitive radios
\cite{Cabric04}\cite{Cabric06} \cite{Haykin2005} via distributed
detection approaches \cite{Blum97}\cite{Varshney97}. However, they
are limited to the detection of signals on a single frequency
band. In \cite{Vistotsky05}, two decision-combining approaches
were studied: hard decision with the AND logic operation and soft
decision using the likelihood ratio test \cite{Blum97}. It was
shown that the soft decision combination of spectrum sensing
results yields gains over hard decision combining. In
\cite{Ghurumuruhan05a}, the authors exploited the fact that
summing signals from two secondary users can increase the
signal-to-noise ratio (SNR) and detection reliability if the
signals are correlated. In \cite{Lundén07}, a generalized
likelihood ratio test for detecting the presence of
cyclostationarity over multiple cyclic frequencies was proposed
and evaluated through Monte Carlo simulations. Along with these
works, we have developed a linear cooperation strategy
\cite{Quan2007c}\cite{Quan2007J} based on the optimal combination
of the local statistics from spatially distributed cognitive
radios. Generally speaking, the quality of the detector depends on
the level of cooperation and the bandwidth of the control channel.

The literature of wideband spectrum sensing for cognitive radio
networks is very limited. An early approach is to use a tunable
narrowband bandpss filter at the RF front-end to sense one narrow
frequency band at a time \cite{Sahai2005}, over which the existing
narrowband spectrum sensing techniques can be applied. In order to
operate over multiple frequency bands at a time, the RF front-end
requires a wideband architecture and the spectrum sensing usually
involves the estimation of the power spectral density (PSD) of the
wideband signal. In \cite{Tian2006} and \cite{Hur2006}, the
wavelet transform was used to estimate the PSD over a wide
frequency range given its multi-resolution features. However, none
of the previous works considers making joint decisions over
multiple frequency bands, which is essential for implementing
efficient cognitive radios networks.

In this paper, we introduce the multiband joint detection
framework for wideband spectrum sensing in individual cognitive
radios. Within this framework, we jointly optimize a bank of
multiple narrowband detectors in order to improve the
opportunistic throughput capacity of cognitive radios and reduce
their interference to the primary communication systems. In
particular, we formulate wideband spectrum sensing into a class of
optimization problems. The objective is to maximize the
opportunistic throughput in an interference limited cognitive
radio network. By exploiting the hidden convexity of the seemingly
non-convex problems, we show that the optimization problems can be
reformulated into convex programs under practical conditions. The
multiband joint detection strategy allows cognitive radios to
efficiently take advantage of the unused frequency bands and limit
the resulting interference.

The rest of this paper is organized as follows. In Section
\ref{sec:model}, we describe the system model for wideband
spectrum sensing. In Section \ref{sec:mb_joint_det}, we develop
the multiband joint detection algorithms, which seek to maximize
the opportunistic throughput. The proposed spectrum sensing
algorithms are examined by numerical examples in Section
\ref{sec:sim} and conclusions are drawn in Section \ref{sec:cls}.

\section{System Models}\label{sec:model}



\subsection{Wideband Spectrum Sensing}

Consider a primary communication system (e.g., a multicarrier
modulation based system) over a wideband channel that is divided
into $K$ non-overlapping narrowband subchannels. In a particular
geographical region and time, some of the $K$ subchannels might
not be utilized by the primary users and are available for
opportunistic spectrum access. Multiuser orthogonal frequency
division multiplexing (OFDM) is an ideal candidate for such a
scenario since it makes the subband manipulation easy and
flexible.

%

\begin{figure}
\centering
\includegraphics[width=3.5in]{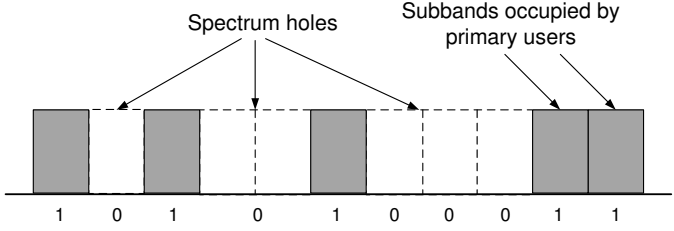}
\caption{A schematic illustration of a multiband channel.}
\label{fig:sppool}
\end{figure}

We model the occupancy detection problem on subchannel $k$ as one
of choosing between $\mathcal{H}_{0,k}$ (``$0$''), which
represents the absence of primary signals, and $\mathcal{H}_{1,k}$
(``$1$''), which represents the presence of primary signals. An
illustrative example where only some of the $K$ bands are occupied
by primary users is depicted in Fig. \ref{fig:sppool}. The
underlying hypothesis vector is a binary representation of the
subchannels that are allowed for or prohibited from opportunistic
spectrum access.

The crucial task of spectrum sensing is to sense the $K$
narrowband subchannels and identify spectral holes for
opportunistic use. For simplicity, we assume that the high-layer
protocols, e.g., the medium access control (MAC) layer, can
guarantee that all cognitive radios keep quiet during the
detection interval such that the only spectral power remaining in
the air is emitted by the primary users in addition to background
noises. In this paper, instead of considering a single subband at
a time, we propose to use a multiband detection technique, which
jointly takes into account the detection of primary users across
multiple frequency bands. We next present the system model.

\subsection{Received Signal}

Consider a multi-path fading environment, where $h(l)$, $l=0, 1,
\ldots, L-1$, denotes the discrete-time channel impulse response
between the primary transmitter and cognitive radio receiver, with
$L$ as the number of resolvable paths. The received baseband
signal at the CR front-end can be represented as
\begin{equation}
r(n)= \sum_{l=0}^{L-1} h\left(l\right)  s\left(n-l\right)  + v(n),
\ \ n=0, 1, \ldots, N-1
\end{equation}
where $s(n)$ is the primary transmitted signal at time $n$ (after
the cyclic prefix has been removed) and $v(n)$ is additive complex
white Gaussian noise with zero mean and variance $\sigma_v^2$,
i.e., $v(n) \sim \mathcal{CN}\left(0, \sigma_v^2 \right)$. In a
multi-path fading environment, the wideband channel exhibits
frequency-selective features
\cite{Proakis2001}\cite{Goldsmith2006} \cite{Sayed2003} and its
discrete frequency response is given by
\begin{equation}
H_k = \frac{1}{\sqrt{N}} \sum_{n=0}^{L-1}  h(n) e^{-j 2 \pi nk/N},
\ \ \ \ \ \ k=0, 1, \ldots, K-1
\end{equation}
where $L \leq N$. We assume that the channel is slowly varying
such that the channel frequency responses $\{H_k\}_{k=0}^{K-1}$
remain constant during a detection interval. In the frequency
domain, the received signal at each subchannel can be estimated by
first computing its
discrete Fourier transform (DFT): 
\begin{align}
R_k & = \frac{1}{\sqrt{N}}\sum_{n=0}^{N-1} r(n)  e^{-j 2\pi n k/N}
\nonumber \\
&= H_k S_k + V_k,  \ \ \ \ \ \ \ \ k=0, 1, \ldots, K-1
\end{align}
where $S_k$ is the primary transmitted signal at subchannel $k$
and
\begin{equation}
V_k = \frac{1}{\sqrt{N}} \sum_{n=0}^{L-1}  v(n) e^{-j 2 \pi nk/N},
\ \ \ \  k=0, 1, \ldots, K-1
\end{equation}
is the received noise in frequency domain. The random variable
$V_k $ is independently and normally distributed with zero mean
and variance $\sigma_v^2$, i.e., $V_k \sim \mathcal{CN}\left(0,
\sigma_v^2 \right)$, since $v(n) \sim \mathcal{CN}\left(0,
\sigma_v^2 \right)$ and the DFT is a linear operation. Without
loss of generality, we assume that the transmitted signal $S_k$,
the channel gain $H_k$, and the additive noise $V_k$ are
independent of each other.

\subsection{Signal Detection in Individual Bands}

Here, we consider signal detection in a single narrowband
subchannel, which will constitute a building block for multiband
joint detection. To decide whether the $k$-th subchannel is
occupied or not, we test the following binary hypotheses
\begin{align}
&\mathcal{H}_{0,k} :~ R_k = V_k \ \ \ \ \ \ \ \ \ \ \ \ \ \ \ \
\nonumber
\\
&\mathcal{H}_{1,k}:~ R_k = H_k S_k +V_k,   \ \ \ \ \ \ k=0, 1,
\ldots, K-1
\end{align}
where $\mathcal{H}_{0,k}$ and $\mathcal{H}_{1,k}$ indicate,
respectively, the absence and presence of the primary signal in
the $k$-th subchannel. For each subchannel $k$, we compute the
summary statistic as the sum of received signal energy over an
interval of $M$ samples, i.e.,
\begin{equation}
Y_k = \sum_{m=0}^{M-1} \left|R_k(m) \right|^2,  \ \ \ \ \ \ k=0,
1, \ldots, K-1
\end{equation}
and the decision rule is given by
\begin{equation} \label{eqn:NarrowBand_Detection}
Y_k  \begin{array}{c} \mathcal{H}_{1,k} \\ \gtreqless \\
\mathcal{H}_{0,k} \end{array}  \gamma_k, \ \ \ \ \ \ \ \ k=0, 1,
\ldots, K-1
\end{equation}
where $\gamma_k$ is the corresponding decision threshold.

For simplicity, we assume that the transmitted signal at each
subchannel has unit power, i.e.,  $\mathbb{E}\left( |S_k|^2
\right) =1$. This assumption holds when primary radios deploy
uniform power transmission strategies given no channel knowledge
at the transmitter side. According to the central limit theorem
\cite{Gendenko1954}, $Y_k$ is asymptotically in $M$ normally
distributed with mean
\begin{equation} \label{eqn:uc_bar}
\mathbb{E} \left(Y_{k}\right) = \left\{
\begin{array}{ll}
    M \sigma_v^2    & \ \ \ \mathcal{H}_{0,k} \vspace{2pt}\\
    M \left(\sigma_v^2 + |H_k|^2 \right)  & \ \ \  \mathcal{H}_{1,k} \\
\end{array}%
\right.
\end{equation}
and variance
\begin{align}
\mathrm{Var} \left(Y_k\right)  = \left\{
\begin{array}{ll}
    2M \sigma_v^4    & \ \ \ \mathcal{H}_{0,k} \vspace{2pt} \\
    2M \left(\sigma_v^2 + 2|H_k|^2 \right)\sigma_v^2   & \ \ \  \mathcal{H}_{1,k} \\
\end{array}%
\right.
\end{align}
for $k=0, 1, \ldots, K-1$. Thus, we write these statistics
compactly as $Y_k \sim \mathcal{N} \left( \mathbb{E} \left(
Y_{k}\right), \mathrm{Var} \left(Y_k\right) \right)$,  $k=0, 1,
\ldots, K-1$.

Using the decision rule in (\ref{eqn:NarrowBand_Detection}), the
probabilities of false alarm and detection at subchannel $k$ can
be respectively calculated as
\begin{align} \label{eqn:P_f}
P_f^{(k)}(\gamma_k) = \mathrm{Pr} \left(Y_k > \gamma_k |
\mathcal{H}_{0,k} \right) = Q \left( \frac{\gamma_k-M \sigma_v^2
}{\sigma_v^2 \sqrt{2M}}\right)
\end{align}
and
\begin{equation} \label{eqn:p_d}
P_d^{(k)}(\gamma_k) = \mathrm{Pr} \left(Y_k > \gamma_k |
\mathcal{H}_{1,k} \right) = Q \left( \frac{\gamma_k-M
\left(\sigma_v^2 + |H_k|^2 \right) }{\sigma_v \sqrt{2M
\left(\sigma_v^2 + 2|H_k|^2 \right)}}\right)
\end{equation}
where $Q(\cdot)$ denotes the complementary distribution function
of the standard normal distribution.

The choice of the threshold $\gamma_k$ leads to a tradeoff between
the probability of false alarm and the probability of
miss\footnote{The subscript $k$ is omitted whenever we refer to a
generic frequency band.}, $P_m=1-P_d$. Specifically, a higher
threshold will result in a smaller probability of false alarm and
a larger probability of miss, and vice versa.

The probabilities of false alarm and miss have unique implications
for cognitive radio networks. Low probabilities of false alarm are
necessary in order to maintain possible high throughput in
cognitive radio systems, since a false alarm would prevent the
unused spectral segments from being accessed by cognitive radios.
On the other hand, the probability of miss measures the
interference from cognitive radios to the primary users, which
should be limited in opportunistic spectrum access. These
implications are based on a typical assumption that if primary
signals are detected, the secondary users should not use the
corresponding channel and that if no primary signals are detected,
then the corresponding frequency band will be occupied by
secondary users.


\begin{figure}
\centering
\includegraphics[width=3.5in]{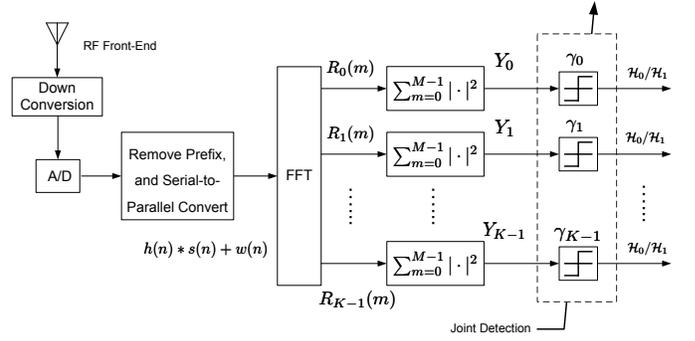}
\caption{A schematic representation of multiband joint detection
for wideband spectrum sensing in cognitive radio networks.}
\label{fig:wss1}
\end{figure}

\section{Multiband Joint Detection} \label{sec:mb_joint_det}

In this section, we present the multiband joint detection
framework for wideband spectrum sensing, as illustrated in Fig.
\ref{fig:wss1}. The design objective is to find the optimal
threshold vector $\boldsymbol{\gamma} = \left[\gamma_0, \gamma_1,
\ldots, \gamma_{K-1} \right]^T$ so that the cognitive radio system
can make efficient use of the unoccupied spectral segments without
causing harmful interference to the primary users. For a given
threshold vector $\boldsymbol{\gamma}$, the probabilities of false
alarm and detection can be compactly represented as
\begin{equation}
\boldsymbol{P}_f(\boldsymbol{\gamma}) = \left[
P_{f}^{(0)}(\gamma_0), P_{f}^{(1)}(\gamma_1), \ldots,
P_{f}^{(K-1)}(\gamma_{K-1}) \right]^T
\end{equation}
and
\begin{equation}
\boldsymbol{P}_d (\boldsymbol{\gamma}) =
\left[P_{d}^{(0)}(\gamma_0), P_{d}^{(1)}(\gamma_1), \ldots,
P_{d}^{(K-1)}(\gamma_{K-1}) \right]^T
\end{equation}
respectively. Similarly, the probabilities of miss can be written
in a vector as
\begin{equation}
\boldsymbol{P}_m (\boldsymbol{\gamma}) =
\left[P_{m}^{(0)}(\gamma_0), P_{m}^{(1)}(\gamma_1), \ldots,
P_{m}^{(K-1)}(\gamma_{K-1}) \right]^T
\end{equation}
where $P_m^{(k)}(\gamma_k) =1- P_d^{(k)}(\gamma_k)  $, $k=0, 1,
\ldots, K-1$, compactly written as $\boldsymbol{P}_m
(\boldsymbol{\gamma}) = \boldsymbol{1} - \boldsymbol{P}_d
(\boldsymbol{\gamma})$, with $\boldsymbol{1}$ the all-one vector.

Consider a cognitive radio sensing the $K$ narrowband subchannels
in order to opportunistically utilize the unused ones for
transmission. Let $r_k$ denote the throughput achievable over the
$k$-th subchannel if used by cognitive radios,
and $\boldsymbol{r}=\left[r_0, r_1, \ldots, r_{K-1}\right]^T$. 
Since $1-P_f^{(k)}$ measures the opportunistic spectrum
utilization of subchannel $k$, we define the aggregate
opportunistic throughput capacity as
\begin{equation}
R\left(\boldsymbol{\gamma}\right) = \boldsymbol{r}^T
\left[\boldsymbol{1}-\boldsymbol{P}_f(\boldsymbol{\gamma}) \right]
\end{equation}
which is a function of the threshold vector $\boldsymbol{\gamma}$.
Due to the inherent trade-off between $P_f^{(k)}(\gamma_k)$ and
$P_m^{(k)}(\gamma_k)$, maximizing the sum rate
$R(\boldsymbol{\gamma})$ will result in large
$\boldsymbol{P}_m(\gamma)$, hence causing harmful interference to
primary users.

The interference to primary users should be limited in a cognitive
radio network. For a widband primary communication system, the
impact of interference induced by cognitive devices can be
characterized by a relative priority vector over the $K$
subchannels, i.e., $\boldsymbol{c}=\left[c_{0}, c_{1}, \ldots,
c_{K-1}\right]^T$, where $c_k$ indicates the cost incurred if the
primary user at subchannel $k$ is interfered with. Suppose that
$J$ primary users share a portion of the $K$ subchannels and each
primary user occupies a subset $S_j$. Consequently, we define the
aggregate interference to primary user $j$ as $ \sum_{i \in S_j}
c_i P_m^{(i)}(\gamma_i) $.
In special cases where each primary user is equally
important, we may have $\boldsymbol{c}=\boldsymbol{1}$.

To summarize, our objective is to find the optimal thresholds
$\{\gamma_k\}_{k=0}^{K-1}$ of these $K$ subchannels, collectively
maximizing the aggregate opportunistic throughput subject to
constraints on the aggregate interference for each primary user
and individual constraints on the subbands. As such, the
optimization problem for a multi-user primary system can be
formulated as
\begin{align} &\mathrm{max}  \ \ \ \ \ R\left(\boldsymbol{\gamma}\right)   \hfill &  \hfill{(\mathrm{P}1)}  \nonumber \\
&  \mathrm{s.t.} \ \ \ \ \ \ \ \sum_{i \in S_j} c_i
P_m^{(i)}(\gamma_i)  \leq \varepsilon_j,\ j=0, 1, \ldots, J-1
\nonumber \\
& \ \ \ \ \ \ \ \ \ \ \ \  \ \ \ \ \ \boldsymbol{P}_m (\boldsymbol{\gamma}) \preceq \boldsymbol{\alpha} \label{eqn:const_Pm} \\
& \ \ \ \ \ \ \ \ \ \ \ \ \ \ \ \ \    \boldsymbol{P}_f
(\boldsymbol{\gamma}) \preceq \boldsymbol{\beta}
\label{eqn:const_Pf}
\end{align}
with the optimization variables
$\boldsymbol{\gamma}=\left[\gamma_0, \gamma_1, \ldots,
\gamma_{K-1} \right]^T$. The constraint (\ref{eqn:const_Pm})
limits the interference on each subchannel with
$\boldsymbol{\alpha} = \left[\alpha_0, \alpha_1, \ldots,
\alpha_{K-1} \right]^T$, and the last constraint in
(\ref{eqn:const_Pf}) dictates that each subchannel should achieve
at least a minimum opportunistic spectrum utilization that is
proportional to $1-\beta_k$. 
For the single-user primary system where all the subchannels are
used by one primary user, we have $J=1$.

Intuitively, we could make some observations on the multiband
joint detection.
First, the subchannel with a higher opportunistic rate $r_k$
should have a higher threshold $\gamma_k$ (i.e., a smaller
probability of false alarm) so that it can be highly used by
cognitive radios. Second, the subchannel that carries a higher
priority primary user should have a lower threshold $\gamma_k$
(i.e., a smaller probability of miss) in order to prevent harmful
interference by secondary users. Third, a little compromise on
those subchannels carrying less important primary users might
boost the aggregate rate considerably. Thus, in the determination
of the optimal threshold vector, it is necessary to strike a
balance among the channel condition, the opportunistic throughput,
and the relative priority of each subchannel.

The objective and constraint functions in ($\mathrm{P}1$) are
generally nonconvex, making it difficult to efficiently solve for
the global optimum. In most cases, suboptimal solutions or
heuristics have to be used. However, we find that this seemingly
nonconvex problem can be made convex by reformulating the problem
and exploiting the hidden convexity.

We observe the fact that the $Q$-function is monotonically
non-increasing allows us to transform the constraints in
(\ref{eqn:const_Pm}) and (\ref{eqn:const_Pf}) into linear
constraints. From (\ref{eqn:const_Pm}), we have
\begin{equation}\label{eqn:trans_const_P_d}
1- P_d^{(k)}(\gamma_k) \leq \alpha_k, \ \ \ \ \ \ k =0, 1, \ldots,
K-1.
\end{equation}
Substituting (\ref{eqn:p_d}) into (\ref{eqn:trans_const_P_d})
gives
\begin{equation}
\gamma_k \leq \gamma_{\mathrm{max}, k} \ \ \ \ \ \ k=0, 1, \ldots,
K-1
\end{equation}
where
\begin{align}
\gamma_{\mathrm{max}, k} \stackrel{\Delta} {=}  M &
\left(\sigma_v^2
+ \left|H_k\right|^2 \right) + \nonumber \\
&\sigma_v \sqrt{2M \left( \sigma_v^2 + 2 \left|H_k\right|^2
\right) } Q^{-1} \left(1-\alpha_k \right).
\end{align}
Similarly, the combination of (\ref{eqn:P_f}) and
(\ref{eqn:const_Pf}) leads to
\begin{equation}
\gamma_k \geq \gamma_{\mathrm{min}, k} \ \ \ \ \ \ k=0, 1, \ldots,
K-1
\end{equation}
where
\begin{equation}
\gamma_{\mathrm{min}, k} = \sigma_v^2 \left[M+ \sqrt{2M} Q^{-1}
\left(\beta_k \right) \right].
\end{equation}
Consequently, the original problem ($\mathrm{P}1$) has the
following equivalent form
\begin{align}
&\mathrm{min}  \ \sum_{k=0}^{K-1} r_k   P_{f}^{(k)}(\gamma_k) \ \ \ \ \ \ \ \ \ \ \ \ \ \ \ \ \ \ \ \ \ \ \ \ \ \  \ \ \ \ \ \ \ \ \ \ \ \ \    (\mathrm{P}2)   \nonumber \\
& \mathrm{s.t.} \ \ \ \  \sum_{i \in S_j} c_i P_m^{(i)}(\gamma_i)
\leq \varepsilon_j, \ j=0, 1, \ldots,
J-1 \label{eqn:P2_cons_1}\\
& \ \ \ \ \ \ \ \ \ \ \ \ \ \gamma_{\mathrm{min}, k} \leq \gamma_k
\leq \gamma_{\mathrm{max}, k}, \ k=0, 1, \ldots, K-1.
\label{eqn:P2_cons_2}
\end{align}
Although the constraint (\ref{eqn:P2_cons_2}) is linear, the
problem is still nonconvex. However, it can be furthermore
transformed into a tractable convex optimization problem in the
regime of low probabilities of false alarm and miss. To establish
the transformation, we need the following results.

\begin{lemma}\label{lemma:1}
The function $P_f^{(k)}\left(\gamma_k \right)$ is convex in
$\gamma_k$ if $P_f^{(k)}\left(\gamma_k \right) \leq \frac{1}{2}$.
\end{lemma}
\begin{proof}
Taking the second derivative of $P_f^{(k)}\left(\gamma_k \right)$
from (\ref{eqn:P_f}) gives
\begin{align}
\frac{d^2 P_f^{(k)}\left(\gamma_k \right)}{d \gamma_k^2}
&=\frac{-1}{\sqrt{2 \pi }} \frac{d}{d \gamma_k }
\exp\left[-\frac{\left(\gamma_k -M \sigma_v^2 \right)^2 }{4 M
\sigma_v^4  } \right] \nonumber \\
& = \frac{\gamma_k -M \sigma_v^2}{2M \sigma_v^2 \sqrt{2 \pi }}
\exp\left[-\frac{\left(\gamma_k -M \sigma_v^2 \right)^2 }{4 M
\sigma_v^4  } \right].
\end{align}
Since $P_f^{(k)}\left(\gamma_k \right) \leq \frac{1}{2}$, we have
$\gamma_k \geq M \sigma_v^2 $. Consequently, the second derivative
of $P_f^{(k)}\left(\gamma_k \right)$ is greater than or equal to
zero, which implies that $P_f^{(k)}\left(\gamma_k \right)$ is
convex in $\gamma_k$.
\end{proof}

\begin{lemma}
The function $P_m^{(k)}\left(\gamma_k \right)$ is convex in
$\gamma_k$ if $P_m^{(k)}\left(\gamma_k \right) \leq \frac{1}{2}$.
\end{lemma}
\begin{proof}
This result can be proved using a similar technique to that used
to prove Lemma \ref{lemma:1}. By taking the second derivative of
(\ref{eqn:p_d}), we can show that $P_d^{(k)}(\gamma_k)$ is
concave, and hence $P_m^{(k)}(\gamma_k)= 1-P_d^{(k)}(\gamma_k)$ is
a convex function.
\end{proof}

Recall that the nonnegative weighted sum of a set of convex
functions is also convex \cite{Boyd2003}. The problem
($\mathrm{P}1$) becomes a convex program if we enforce the
following conditions:
\begin{equation}\label{eqn:pratical_conditions}
0 < \alpha_k \leq \frac{1}{2} \ \ \mathrm{and} \ \ 0< \beta_k \leq
\frac{1}{2}, \ \ \ k=0, 1, 2, \ldots, K-1.
\end{equation}
This regime of probabilities of false alarm and miss is that of
practical interest in cognitive radio networks.



With the conditions in (\ref{eqn:pratical_conditions}), the
feasible set of problem ($\mathrm{P}2$) is convex. The
optimization problem takes the form of minimizing a convex
function subject to a convex constraint, and thus a local maximum
is also the global maximum. Efficient numerical search algorithms
such as the interior-point method can be used to solve for the
optimal solutions \cite{Boyd2003}.

Alternatively, we can formulate the multiband joint detection
problem into another optimization problem that minimizes the
interference from cognitive radios to the primary communication
system, subject to some constraints on the aggregate opportunistic
throughput, i.e.,
\begin{align} &\mathrm{minimize}  \ \ \ \ \ \boldsymbol{c}^T  \boldsymbol{P}_m
(\boldsymbol{\gamma})  \hfill &  \hfill{(\mathrm{P}3)}  \nonumber \\
&\ \ \ \ \ \  \mathrm{st.} \ \ \ \ \ \ \     \boldsymbol{r}^T
\left[\boldsymbol{1}-\boldsymbol{P}_f(\boldsymbol{\gamma}) \right]
 \geq \delta
\nonumber \\
& \ \ \ \ \ \ \ \ \ \ \ \ \ \ \ \   \boldsymbol{P}_m (\boldsymbol{\gamma}) \preceq \boldsymbol{\alpha} \nonumber \\
& \ \ \ \ \ \ \ \ \ \ \ \  \ \ \ \  \boldsymbol{P}_f
(\boldsymbol{\gamma}) \preceq \boldsymbol{\beta} \nonumber
\end{align}
with $\delta$ the required minimum aggregated rate and
$\boldsymbol{\gamma}$ the optimization variables. Like problem
($\mathrm{P}1$), this problem can be transformed into a convex
optimization problem by enforcing the conditions in
(\ref{eqn:pratical_conditions}). The result will be illustrated
numerically later in Section \ref{sec:sim}.

\section{Simulation Results}\label{sec:sim}


In this section, we numerically evaluate the proposed spectrum
sensing schemes. Consider a multiband single-user OFDM system in
which a wideband channel is equally divided into $8$ subchannels.
Each subchannel has a channel gain $H_k$ between the primary user
and the cognitive radio, a throughput rate $r_k$ if used by
cognitive radios, and a cost coefficient $c_k$ indicating a
penalty incurred when the primary signal is interfered with by the
cognitive radio. For each subchannel $k$ ($0 \leq k \leq 7$), it
is expected that the opportunistic spectrum utilization is at
least $50\%$, i.e., $\beta_k = 0.5$, and the probability that the
primary user is interfered with is at most $\alpha_k=0.1$. For
simplicity, it is assumed that the noise power level is
$\sigma_v^2=1$ and the length of each detection interval is
$M=100$. This example studies multiband joint detection in a
single cognitive radio. The proposed spectrum sensing algorithms
are examined by comparing with an approach that searches a uniform
threshold to maximize the aggregate opportunistic throughput. We
randomly generate the channel condition between the primary user
and the cognitive radio, the opportunistic throughput over each
subchannel, and the cost of interference of each subchannel. One
realization example is given in Table \ref{table:ex1}.

\begin{figure}
\centering
\includegraphics[width=3.4in]{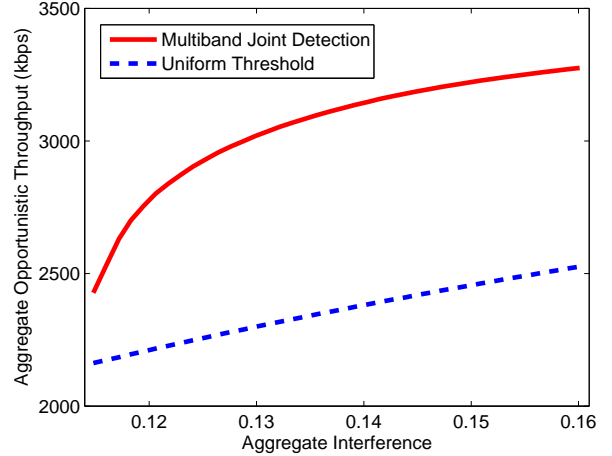}
\caption{The aggregate opportunistic throughput capacity vs. the
constraint on the aggregate interference to the primary
communication system.} \label{fig:Th_vs_Int}
\end{figure}

We maximize the aggregate opportunistic throughput over the $8$
subchannels subject to some constraints on the interference to the
primary users, as formulated in ($\mathrm{P}1$). Fig.
\ref{fig:Th_vs_Int} plots the maximum aggregate opportunistic
rates against the aggregate interference to the primary
communication system. It can be seen that the multiband joint
detection algorithm with optimized thresholds can achieve a much
higher opportunistic rate than that achieved by the one with
uniform threshold. Note that in the reference algorithm, the
uniform threshold is searched to maximize the achievable rate for
a fair comparison. That is, the proposed multiband joint detection
algorithm makes better use of the wide spectrum by balancing the
conflict between improving spectrum utilization and reducing the
interference. In addition, it is observed that the aggregate
opportunistic rate increases as we relax the constraint on the
aggregate interference $\varepsilon$.

\begin{table}
\renewcommand{\arraystretch}{1.3}
\caption{Parameters Used in Simulations} \label{table:ex1}
\centering
\begin{tabular}{c||c|c|c|c|c|c|c|c}
\hline
\bfseries $|H_k|^2$ &  .50 &  .30 &  .45 &  .65 &  .25 &  .60 &  .40 &  .70 \\
\hline
$\boldsymbol{r}$ (kbps) &  612 &  524 & 623 & 139 & 451 & 409 & 909 & 401  \\
\hline
$\boldsymbol{c}$ & 1.91 &  8.17 &  4.23  & 3.86 &  7.16 &  6.05 &  0.82 & 1.30  \\
\hline
\end{tabular}
\end{table}

\begin{figure}
\centering
\includegraphics[width=3.4in]{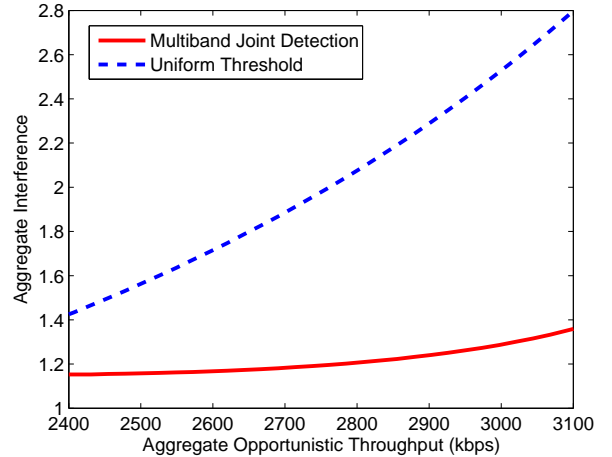}
\caption{The aggregate interference to the primary communication
system vs. the constraint on the aggregate opportunistic
throughput.} \label{fig:Int_vs_Th} \vspace{-10pt}
\end{figure}

\begin{figure}
\centering
\includegraphics[width=3.5in]{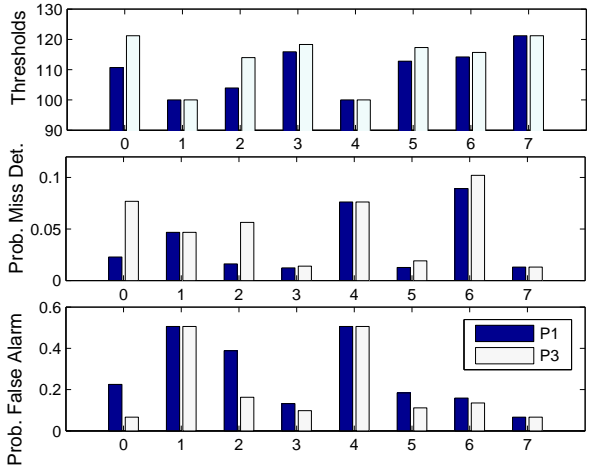}
\caption{The optimized thresholds and the associated probabilities
of miss and false alarm: ($\mathrm{P}1$)  $\varepsilon = 1.25$ and
($\mathrm{P}3$) $\delta = 3224$ kbps.}
 \label{fig:Bar_Int_vs_Th}
\end{figure}

An alternative example is depicted in Fig. \ref{fig:Int_vs_Th},
showing the numerical results of minimizing the aggregate
interference subject to the constraints on the opportunistic
throughput as formulated in ($\mathrm{P}3$). It can be observed
that the multiband joint detection strategy outperforms the one
using uniform thresholds in terms of the induced interference to
the primary users for any given opportunistic throughput. For
illustration purposes, the optimized thresholds and the associated
probabilities of miss and false alarm are given in Fig.
\ref{fig:Bar_Int_vs_Th} for ($\mathrm{P}1$) and ($\mathrm{P}3$).
To summarize, these numerical results show that multiband joint
detection can considerably improve the spectrum efficiency by
making more efficient use of the spectral diversity.

\section{Conclusion}\label{sec:cls}

In this paper, we have proposed a multiband joint detection
approach for wideband spectrum sensing in cognitive radio
networks. The basic strategy is to take into account the detection
of primary users across a bank of narrowband subchannels jointly
rather than to consider only one single band at a time. We have
formulated the joint detection problem into a class of
optimization problems to improve the spectral efficiency and
reduce the interference. By exploiting the hidden convexity in the
seemingly nonconvex problems, we have obtained the optimal
solution under practical conditions. The proposed spectrum sensing
algorithms have been examined numerically and shown to be able to
perform well.

\section*{Acknowledgment}

This research was supported in part by the National Science
Foundation under Grants ANI-03-38807, CNS-06-25637, ECS-0601266,
ECS-0725441, CNS-0721935, CCF-0726740, and by the Department of
Defense under Grant HDTRA-07-1-0037.


\bibliographystyle{IEEEtran}
\bibliography{IEEEabrv,ref}

\end{document}